\documentclass[useAMS,usenatbib,letters]{mn2e}
\usepackage{graphicx}

\newcommand{\mdot}{\dot{m}}
\newcommand{\apj}{ApJ}
\newcommand{\apjl}{ApJL}
\newcommand{\mnras}{MNRAS}
\newcommand{\aap}{A\&A}
\newcommand{\araa}{ARA\&A}
\newcommand{\nat}{Nature}
\newcommand{\apss}{Ap\&SS}

\author[P. O. Petrucci et al.]{Pierre-Olivier Petrucci\thanks{E-mail:
pierre-olivier.petrucci@obs.ujf-grenoble.fr}, Jonathan Ferreira, Gilles Henri and Guy Pelletier\\
Laboratoire d'Astrophysique, Observatoire de Grenoble BP53,
  F-38041 Grenoble cedex 9, France} 

\begin{document}

\date{Received .../Accepted ...}

\title[Disk magnetization and hysteresis in XRBs]{The role of the disk magnetization on the hysteresis behavior of X-ray binaries}

\maketitle

\begin{abstract}
We present a framework for understanding the dynamical and spectral properties of X-ray Binaries, where the presence of an organized large scale magnetic field plays a major role.  Such a field is threading the whole accretion disk with an amplitude measured by the disk magnetization $\mu(r,t) =B_z^2/(\mu_o P_{tot})$, where $P_{tot}$ is the total, gas and radiation, pressure. 
Below a transition radius $r_J$, a jet emitting disk (the JED) is settled and drives self-collimated non relativistic jets. Beyond $r_J$, no jet is produced despite the presence of the magnetic field and a standard accretion disc (the SAD) is established. The radial distribution of the disk magnetization $\mu$ adjusts itself to any change of the disk accretion rate $\dot m$, thereby modifying the transition radius $r_J$. 

We propose that a SAD-to-JED transition occurs locally, at a given radius, in a SAD when $\mu=\mu_{max} \simeq 1$ while the reverse transition occurs in a JED only when $\mu=\mu_{min}\simeq 0.1$. 
This bimodal behavior of the accretion disk provides a promising way to explain the hysteresis cycles followed by  X-ray binaries during outbursts.

\end{abstract}


\section{Introduction}

In the last ten years, different studies have shown that during periods of strong activity XRBs follow an  hysteresis cycle  in the hardness--intensity diagram (hereafter HID), the X-ray equivalent of the color-magnitude diagrams used in the optical range \citep{miy95,now02,bar02,rod03,mac03c,cor04,bel05,zdz04b,rem06}.  The archetype of HID has been reported in Fig. \ref{muprofil} in the case of GX 339-4 during its 2002/2003 outburst. The system transits from radio loud (jet dominated) hard states (HS) at the right in the figure to radio quiet (disk dominated) soft states (SS) at the left. Such transitions occur at a luminosity generally a factor of a few larger than during  the SS-HS transitions back to the initial state. This hysteresis cycle clearly indicates that the accretion rate cannot be the unique physical parameter that controls the spectral evolution of these systems. And the fact that at some point XRBs produce collimated jets (as probed by flat radio spectra) points towards the other physical parameter: the magnetic field.

In a previous paper (\citealt{fer06a} hereafter F06), we presented a new model to explain the different spectral states observed in Black Hole XRB that assumes the presence of  a large
scale vertical magnetic field $B_z$ (of the same polarity everywhere) anchored in an accretion disk. The central regions have a multi-flow configuration consisting of an outer standard accretion disk (SAD) down to a transition radius $r_{J}$ and an inner magnetized jet emitting disk (JED)  below $r_{J}$. The JED drives  a self-collimated non relativistic MHD jet surrounding, when
conditions for pair creation are met,  a ultra relativistic pair beam on its axis. We also assume the presence of a hot thermal corona at the base of the jet that is likely part of the JED itself (\citealt{pet06b}, Petrucci et al. 2007 in prep.). This model provides a promising framework to explain the canonical spectral states of BH XRBs mainly by varying $r_{J}$ and $\dot{m}$\footnote{$\dot{m}=\dot{M}/\dot{M}_{edd}$ with $\dot{M}$ the accretion rate and $\dot{M}_{edd}=L_{edd}/c^2$ with $L_{edd}$ the Eddington luminosity and $c$ the speed of light.} independently. The hard states then correspond to large  $r_{J}$ while soft ones correspond to small $r_{J}$. This transition radius is observationally identified as the usual "disk inner radius $r_{in}$" determined from spectral fits. Indeed, the JED is transferring most of the released accretion power into the jets: the black body spectrum is therefore dominated by the outer SAD.

However this model did not address the cause of the transitions between the various spectral states, namely what really drives the variations of $r_{J}$. This directly translates to the fundamental question: what are the physical conditions to have a JED i.e. to produce a large scale MHD jet from an accretion disk? This points out to the important local parameter called the disk magnetization 
\begin{equation}
\mu(r,t) = \frac{ B_z^2}{\mu_o P_{tot}}
\end{equation} 
where $P_{tot}= P_{gas} + P_{rad}$ includes the plasma and radiation pressures\footnote{Note that $\mu$ differs from the inverse of the usual plasma $\beta$ parameter which compares the magnetic pressure to the gas pressure only.}.  The disk magnetization is a fundamental parameter in accretion-ejection physics and, as shown analytically by \cite{fer95} and \cite{fer97}, and numerically by e.g. \cite{cas02} and \cite{zan07}, it has to be close to unity (i.e. the magnetic field must be close to equipartition) to produce  super-Alfv\'enic jets. We here address the crucial role of the disk magnetization in the spectral evolution of XRBs. We show that its expected variation during the outburst, coupled with that of the accretion rate,  can explain the hysteresis cycle  across the HID.

\section{The disk magnetization}
\label{diskmag}

The reason why the disk magnetization has to be close to unity to produce steady-state jets is
twofold. On one hand, the magnetic field is vertically pinching the
accretion disk so that a (quasi) vertical equilibrium is obtained only
thanks to the gas and radiation pressure support. As a consequence, the
field cannot be too strong. But on the other hand, the field must be
strong enough to accelerate efficiently the plasma right at the disk
surface so that the slow-magnetosonic point is crossed smoothly. These
two constraints can only be met with fields close to equipartition. Quantitatively, analytical self-similar calculations show that $\mu$ has to lie between $\mu_{min}\simeq 0.1$ and $\mu_{max}\simeq 1$ for a stationnary jet from a keplerian disk to exist (\citealt{fer95}, see e.g. Fig.~2 in \citealt{fer97}). The existence of a limited range in $\mu$ is a key point in our interpretation of the hysteresis cycle.

\begin{figure*}
\centering
\includegraphics[width=0.74\textwidth]{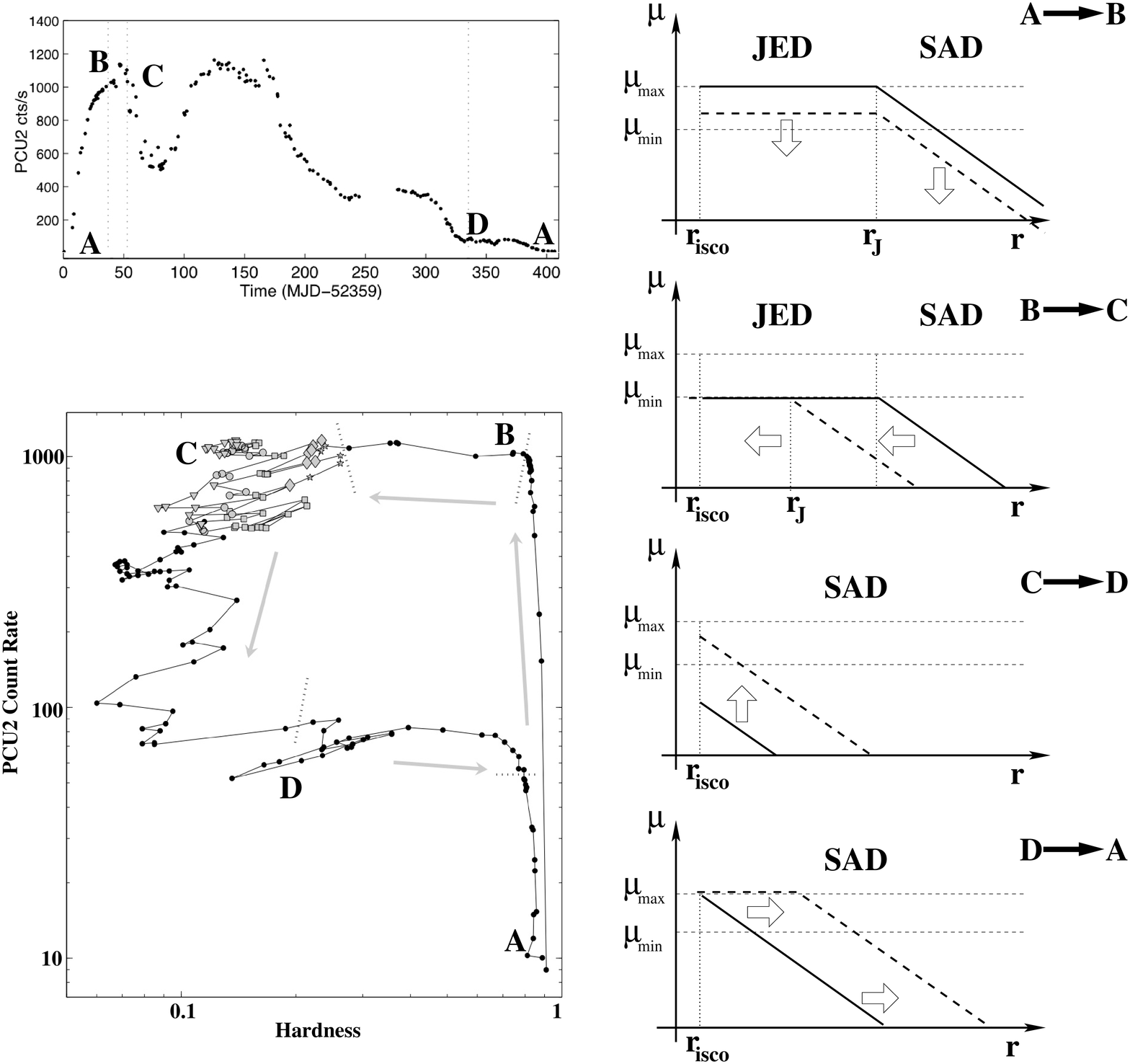}
\caption{{\bf Left:} Light curve and Hardness-Intensity Diagram followed by GX 339-4 during its 2002-2003 outburst (from \citealt{bel05}). {\bf Right:} Sketch of the expected evolution of the radial distribution of the disk magnetization $\mu(r)$ during an outburst with hysteresis. In each plot, the solid line represents the initial (for each part of the HID) radial distribution of the disk magnetization, while the dashed one shows its evolution. See Sect. \ref{evolution} for more details. \label{muprofil}}
\end{figure*}

In our framework, a large scale, organized vertical magnetic field $B_z$ is threading the whole accretion disk.  There have been some claims in the literature that such a field could not be maintained in a SAD (see e.g. \citealt{lub94}). In fact, the magnetic field distribution is given by the interplay between advection and diffusion, namely Ohm's law. It turns out that in a SAD in a stationary state, this equation gives $B_z \propto r^{-\mathcal{R}_m}$ with $\mathcal{R}_m\simeq\mathcal{R}_e\simeq 1$. As a consequence, one can reasonably expect  $\mu(r)$ to be a decreasing function of the radius (cf. Eq. 4 of F06). On the other hand, in a JED one has roughly $B_z \propto r^{-5/4}$ \citep{fer95} and $\mu$ remains constant with the radius. In consequence, a reasonable radial distribution of $\mu$ should look like the one plotted on top of Fig. \ref{muprofil}, namely constant (but in the range $\mu_{min}$--$\mu_{max}$)  in the JED, located between the disk innermost stable circular orbit  $r_{ISCO}$ and $r_{J}$, and decreasing in the SAD beyond $r_{J}$. 

Now, the evolution in time of the disk magnetization $\mu(r,t)$ is a far more complex issue. Since $P_{tot} \propto \dot m$, one gets $\mu \propto \dot m^{-1} B_z^2$. It means that its temporal evolution is governed by the evolution of both  $B_z$ and $\dot m(t)$. Although the origin of the variations of the disk accretion rate $\dot m(t)$ during an outburst is relatively well understood (e.g. \citealt{las01}), the evolution of $B_z(t)$ is more speculative. One of the reasons is that $B_z$ could itself be a function of the disk accretion rate. Different scenarios have been proposed (e.g. \citealt{tag04,kin04}) but the problem is very tricky and no detailed calculation has emerged so far. In this context, and given the large uncertainties on the physical processes governing the evolution of the magnetic field with $\dot m$, we make the zeroth order approximation that  the variation of the accretion rate is the dominant cause of the variation of $\mu$. In other words we assume here that $\mu(r,t)$ anti-correlates with the accretion rate. 

One last issue is what determines the transition radius $r_J$ in a disk that is threaded everywhere by a large scale $B_z$ field? In a SAD, the origin of the turbulence is commonly believed to arise from the magneto-rotational instability or MRI \citep{bal91,les07}. However the MRI requires the presence of a weak magnetic field and is quenched when the field is at equipartition. Thus, while a SAD works well at $\mu \ll 1$, there is no known solution when $\mu \ga 1$. On the other hand, as said before, a steady-state JED can establish only around a rather limited range in disk magnetization, namely $\mu_{min} \la \mu \la \mu_{max}$. 

These facts lead us to make the following conjecture that, at a given radius, 

- a SAD-to-JED transition occurs when $\mu=\mu_{max}\simeq 1$,

- a JED-to-SAD transition occurs when $\mu =\mu_{min}\simeq 0.1$. 

As we will see next Section, this conjecture provides a plausible scenario for the hysteresis cycle observed in BH XRBs.

\section{Temporal evolution of BH XRBs}
\label{evolution}

It is commonly believed that BH XRBs generally evolve along an hysteresis cycle during their outbursts, following a counter clockwise sequence A-B-C-D like the one reported in Fig.~\ref{muprofil}, before turning back to their initial state at the end of the outburst. The HS and SS  correspond to the A-B and C-D parts of this diagram respectively. The horizontal branches, B-C and D-A, correspond to the  intermediate states (hereafter IS) whose spectral and timing properties are relatively complex (e.g. \citealt{bel05,fen04,cor04}). This is especially the case during the high IS (hereafter HIS, \citealt{hom05,bel05}) that corresponds to the top horizontal branch where the system luminosity is high.  During its evolution along this horizontal branch the system may enter in a flaring state associated with radio and X-ray flares and/or superluminal sporadic ejections (e.g. \citealt{sob00,han01,fen04c}).  This spectral evolution goes also with a quenching of the radio emission, signature of the disappearance of the jet component. During the bottom horizontal branch (the low IS, hereafter LIS) and the turning back of the system to its initial state, the radio emission is detected again, meaning that the ejection structure has been rebuilt. 

We detail below our interpretation of this hysteresis cycle as a consequence of our conjecture on the disk magnetization $\mu(r,t)$. We give also some estimates of the expected time scales involved at the different phases of the cycle. In the accretion disk, the shortest dynamical time scale dealt with is the keplerian orbit time, namely $\tau_D (r) =  2\pi/\Omega_k = 0.1\ (m/10)(r/50 r_{g})^{3/2}$ sec, where $m$ is the black hole mass in solar mass units and $r_{g}=GM/c^2$ the gravitational radius. In the following we take $m=10$ and consequently $r_g=1.5\times 10^6$ cm.

Let us start at a Low/hard State located at the bottom of the HID right branch (point A in Fig.~\ref{muprofil}). This state is usually seen as a SAD being interrupted at typically  $r_{in} \sim 100 r_{g}$. Within our framework, this is perfectly consistent with a JED settled from $r_{ISCO}$ up to $r_{J} = r_{in}$ and driving self-collimated non relativistic jets (seen as persistent radio features). Within the JED $\mu$ is roughly constant  and we assume that at the beginning of the outburst $\mu = \mu_{max}\simeq 1$. Let us now consider an increase in the disk accretion rate $\dot m$ (an outburst) triggered by, for instance, a disk instability at some outer radius $r_{out}\gg r_{J}$ (e.g. \citealt{las01}). A front defined by a rise in $P_{tot}$ (hence a decrease in $\mu$) propagates then radially inwards eventually reaching $r_J$. As for the SAD, the JED adapts itself locally on a dynamical time scale (a few $\tau_D$) by lowering $\mu$. This adaptation will affect the whole JED extension on a timescale roughly equal to $\tau_{acc,JED}$, the accretion time scale in the JED.  Since the accretion radial velocity  $u_r^{JED} \sim \Omega_K h$ in a JED \citep{fer97}, this time is even shorter than that of the front propagation in the outer SAD. For instance, one gets $\tau_{acc,JED}(100 r_g) \simeq \tau_D(100 r_g)r/h \simeq 30$ sec (using a disk aspect ratio  $h/r=0.01$, \citealt{sha73}). This timescale is much shorter than the timescale of the ascendent phase of an outburst, which is typically of the order of several days between states A and B. Thus, during the ascendant motion of the XRB along the right branch from A to B, $\mu$ should only decrease within the whole JED without any change in $r_{J}$: no spectral state transition is expected.
 
If the rise in $\dot m$ is such that $\mu(r_{J})$ decreases below the critical value $\mu_{min}$, then the XRB arrives at point B. Any subsequent increase of the accretion rate will then produce an outside-in JED-to-SAD transition starting from the outer part of the JED. The transition radius $r_{J}$ decreases and both the MHD jet and the corona extension shrink. The XRB enters the high intermediate state, between B and C. Since $r_{J}$ is decreasing, the timescale for the system evolution along the top branch becomes more and more rapid, in agreement with observations (e.g. \citealt{bel05}).

Formally the end of the top branch should be reached  around point C when $r_{J}=r_{ISCO}$ i.e. when the JED has completely disappeared and the whole disk adopts a radial structure akin to the standard disk model.  However, we expect this situation to be quite variable, probably even before $r_{J}=r_{ISCO}$. Indeed, an important large scale (open) magnetic field is present, providing a magnetic torque helping the MRI to drive accretion. But it is not clear whether jets can be formed or not. In fact, the only calculations that actually proved that steady ejection can take place from a disk were done with a self-similar ansatz. Implicitly, this assumes that the radial extension of the jet emitting region is large. This is consistent with the physical interpretation that the disk material must be first radially accreting fast before being lifted up by the plasma pressure gradient and magneto-centrifugally accelerated at the disk surface by the Lorentz force. What will happen when $r_{J} \ga r_{ISCO}$ only? Our guess is that the MHD jet structure will probably be oscillating between life and death. Accordingly, persistent radio jets should thus disappear even before point C is reached. Besides, in our framework, we also expect  luminous systems to enter a flaring state as soon as the conditions for a strong  pair production in the MHD funnel are fulfilled\footnote{This would correspond to the jet line defined by \cite{fen04c}, see F06 for more details.}, resulting in sporadic ultra-relativistic blob ejections. This could explain why this HID region seems to harbor complex variability phenomena (e.g. \citealt{bel05,nesp03}). 

In C, the system enters the so-called Soft State. The disappearance of the MHD jet would explain the quenching of the radio emission generally observed at this stage (e.g. \citealt{cor03,fen04c}). Then whatever the accretion rate does, and following our conjecture, the JED cannot reappear along the left vertical branch unless $\mu(r_{ISCO})$ reaches $\mu_{max} \sim 1$. The descent from C to D must then correspond to a decrease in intensity i.e. mainly to a decrease of the accretion rate itself, with no spectral state transition. This is the beginning of the fading phase of the outburst. In our simple picture, the decrease of the disk accretion rate produces an outside-in decrease in $P_{tot}$ leading to an increase in $\mu$. Point D then corresponds to the situation when $\mu(r_{ISCO})=\mu_{max}$. 

This initiates the inside-out rebuilding of the accretion-ejection structure along the low horizontal branch. Note that the accretion rate must still be decreasing. But it
does not require a strong decrease in $\dot m$ to build up again a JED on a significant extension (hence jets observable in radio and a change in the hardness ratio). Indeed, as said before, the steady-state radial distribution of the vertical magnetic field is steeper in a SAD
($B_z \propto r^{-{\cal R}_m}$, with ${\cal R}_m \simeq 1$) than in a JED ($B_z \propto
r^{-5/4}$). As a consequence, $\mu$ may become greater than unity in the
innermost disk regions with only a slight decrease in $\dot m$. There is no known steady state disk solution at these large values of magnetization. We expect such a strong field to diffuse away outwardly in the outer SAD so that a maximum value of $\mu =1$ can be maintained in the inner JED-like regions. Providing a time scale is out of the scope of the present paper as it would require to take into account the whole process of the magnetic field redistribution. However, this time is probably shorter than (in the JED) or comparable to (in the SAD) the accretion time scale. We thus expect a rather fast inside-out rebuilding of the JED with reappearance of the self-collimated electron-proton jet and the hot corona. This marks also a  spectral state transition as the SAD emission is receding when going back to point A,  at the end of the outburst.

\section{Discussion and Conclusion} 
\label{discussion}

We present a scenario where the spectral states of X-ray binaries depend on the transition radius $r_{J}$ between an inner accretion-ejection structure and an outer standard accretion disk (F06). We propose in this paper that the variation  of $r_{J}$ is  mostly controlled by the evolution of the accretion rate $\dot m$ and the disk magnetization $\mu$. 

We made the conjecture that a SAD-to-JED transition will locally occur when $\mu$ reaches a maximum value $\mu_{max} \simeq 1$, whereas a JED-to-SAD transition requires a much lower $\mu$ : analytical estimates show that it could happen around $\mu_{min} \simeq 0.1$. It was shown that such a simple framework is enough to explain the dynamical and spectral behaviors of XRBs, providing moreover a dynamical basis for the hysteresis cycles observed in the HIDs. The exact values taken for the limits $\mu_{min}$ and $\mu_{max}$ may not be strictly equal to those chosen here as they were derived from self-similar calculations \citep{fer97}. However, this is not crucial as the main element is the existence of such an interval. It is noteworthy though that these values are compatible with our current understanding of disk physics: there are neither SAD solutions with large scale fields beyond equipartition, nor JED solutions at much lower fields. However, a precise examination of the dynamical transitions between a JED to a SAD solution and vice versa remains to be done in order to assess the conjecture made here.

Interestingly this scenario gives a physical  constraint on the existence of a hysteresis cycle in a given source during an outburst. Indeed, the magnetization has to decrease necessarily below $\mu_{min}$ {\it in the whole accretion disk} during the flare. This translates into the condition that, at a given time during the burst, $\mu (r_{ISCO})$ has to become smaller than $\mu_{min}$. If it is not the case the source should oscillate between hard and soft states with an inner JED never disappearing. Cyg X-1 could be in this situation. It is known to transit from hard to soft state and reversely at the same accretion rate $\dot m \sim 0.01$ (e.g. \citealt{zdz02b,smi02}). It is persistent in X but also in radio, even if the radio emission is weaker in its softest states (\citealt{brock99} and references therein, \citealt{pan06}). This suggests that ejection is always present. On the other hand the strong disk signature in the soft state implies that $r_{J}$ is close to $r_{ISCO}$. Thus for some reasons, in this object  the accretion-ejection structure keeps stable even for low value of $r_{J}$. That could be linked to the fact that it is a high mass X-ray binary system as already proposed by \citet{smi02,mac03c}. 

GRS 1915+105 has also a peculiar behavior that does not match the one described in Sect. \ref{evolution}. This source is generally believed to accrete close to its Eddington limit and to stay preferentially at the top of the HID. Its "hard"  states would correspond to the right part of the top branch while its soft states are generally observed in the left part of the top branch, after the so-called jet line \citep{fen04c}. The flare states are observed when the source oscillates around this line. In our view, in this region of the HID the JED competes with the SAD with $r_{J}\sim r_{ISCO}$. The magnetic field keeps trying to re-adjust itself in the inner disk region to agree with one of the two disk solutions. This might produce the complex variability pattern exhibited by this source during outbursts \citep{bel00,fen04b}. Moreover, close to the ISCO the time scales controlling the system are of the order of fractions of second, in agreement with the rapid variability that characterizes this peculiar object.

Maccarone \& Coppi (2003) and \citet{zdz04} provided several other suggestions for the hysteresis behaviour of XRBs but without going much into details (see critical discussion in \citealt{mey05}). In fact, it seems that the most detailed model so far is the disk evaporation model (\citealt{mey05} and references therein). In this model the efficiency of the SAD evaporation critically depends on the external photon field. In the Soft State for instance, there is mainly a Compton cooling and not much evaporation.  The spectral state transitions would then be triggered at different accretion rates simply because of history, hence of a different Comptonizing spectrum (see their Fig.~2). Although the physics of jet production is not included in this model, taking into account radiative effects in order to explain the HID hysteresis sounds quite reasonable and could, in principle, be integrated in our framework.  In our case however the hysteresis has a dynamical explanation based on the disk magnetization and a two-temperature disk never settles in. Since ejection processes are known to be important in the XRB global energy budget, our suspicion is therefore that evaporation is not the main driver.

As a final note, we stress that if one is to explain self-collimated jets in XRBs, then large scale magnetic fields in the inner regions of accretion disks must be taken into account. Their presence is both compatible with standard accretion disk physics and provides a simple and promising framework to explain hysteresis cycles. In our view, the magnetic flux available in disks is therefore a fundamental and unavoidable ingredient for modeling XRBs. Also, there is no reason why it should not vary from one system to another. Since changing the amount of magnetic flux changes the transition radius $r_{J}$, the characteristic value of  $\mdot$ (hence luminosity) associated with each spectral state transition is also modified. This picture becomes even more complex if accreting material is advecting magnetic flux from the secondary. Indeed, taking into account the advection of a large scale magnetic fields within the disk introduces a whole new set of variable phenomena.


\begin{thebibliography}{}

\bibitem[\protect\citeauthoryear{{Balbus} \& {Hawley}}{{Balbus} \&
  {Hawley}}{1991}]{bal91}
{Balbus} S.~A.,  {Hawley} J.~F.,  1991, \apj, 376, 214

\bibitem[\protect\citeauthoryear{{Barret} \& {Olive}}{{Barret} \&
  {Olive}}{2002}]{bar02}
{Barret} D.,  {Olive} J.,  2002, \apj, 576, 391

\bibitem[\protect\citeauthoryear{{Belloni}, {Homan}, {Casella}, {van der Klis},
  {Nespoli}, {Lewin}, {Miller} \& {Mendez}}{{Belloni} et~al.}{2005}]{bel05}
{Belloni} T.,  {Homan} J.,  {Casella} P.,  {van der Klis} M.,  {Nespoli} E.,
  {Lewin} W.,  {Miller} J.,    {Mendez} M.,  2005, astro-ph/0504577

\bibitem[\protect\citeauthoryear{{Belloni}, {Klein-Wolt}, {M{\'e}ndez}, {van
  der Klis} \& {van Paradijs}}{{Belloni} et~al.}{2000}]{bel00}
{Belloni} T.,  {Klein-Wolt} M.,  {M{\'e}ndez} M.,  {van der Klis} M.,    {van
  Paradijs} J.,  2000, \aap, 355, 271

\bibitem[\protect\citeauthoryear{{Brocksopp}, {Fender}, {Larionov}, {Lyuty},
  {Tarasov}, {Pooley}, {Paciesas} \& {Roche}}{{Brocksopp}
  et~al.}{1999}]{brock99}
{Brocksopp} C.,  {Fender} R.~P.,  {Larionov} V.,  {Lyuty} V.~M.,  {Tarasov}
  A.~E.,  {Pooley} G.~G.,  {Paciesas} W.~S.,    {Roche} P.,  1999, \mnras, 309,
  1063

\bibitem[\protect\citeauthoryear{{Casse} \& {Keppens}}{{Casse} \&
  {Keppens}}{2002}]{cas02}
{Casse} F.,  {Keppens} R.,  2002, \apj, 581, 988

\bibitem[\protect\citeauthoryear{{Corbel}, {Fender}, {Tomsick}, {Tzioumis} \&
  {Tingay}}{{Corbel} et~al.}{2004}]{cor04}
{Corbel} S.,  {Fender} R.~P.,  {Tomsick} J.~A.,  {Tzioumis} A.~K.,    {Tingay}
  S.,  2004, \apj, 617, 1272

\bibitem[\protect\citeauthoryear{{Corbel}, {Nowak}, {Fender}, {Tzioumis} \&
  {Markoff}}{{Corbel} et~al.}{2003}]{cor03}
{Corbel} S.,  {Nowak} M.~A.,  {Fender} R.~P.,  {Tzioumis} A.~K.,    {Markoff}
  S.,  2003, \aap, 400, 1007

\bibitem[\protect\citeauthoryear{{Fender} \& {Belloni}}{{Fender} \&
  {Belloni}}{2004}]{fen04b}
{Fender} R.,  {Belloni} T.,  2004, \araa, 42, 317

\bibitem[\protect\citeauthoryear{{Fender}, {Wu}, {Johnston}, {Tzioumis},
  {Jonker}, {Spencer} \& {van der Klis}}{{Fender} et~al.}{2004}]{fen04}
{Fender} R.,  {Wu} K.,  {Johnston} H.,  {Tzioumis} T.,  {Jonker} P.,  {Spencer}
  R.,    {van der Klis} M.,  2004, \nat, 427, 222

\bibitem[\protect\citeauthoryear{{Fender}, {Belloni} \& {Gallo}}{{Fender}
  et~al.}{2004}]{fen04c}
{Fender} R.~P.,  {Belloni} T.~M.,    {Gallo} E.,  2004, \mnras, 355, 1105

\bibitem[\protect\citeauthoryear{{Ferreira}}{{Ferreira}}{1997}]{fer97}
{Ferreira} J.,  1997, \aap, 319, 340

\bibitem[\protect\citeauthoryear{{Ferreira} \& {Pelletier}}{{Ferreira} \&
  {Pelletier}}{1995}]{fer95}
{Ferreira} J.,  {Pelletier} G.,  1995, \aap, 295, 807+

\bibitem[\protect\citeauthoryear{{Ferreira}, {Petrucci}, {Henri}, {Saug{\'e}}
  \& {Pelletier}}{{Ferreira} et~al.}{2006}]{fer06a}
{Ferreira} J.,  {Petrucci} P.-O.,  {Henri} G.,  {Saug{\'e}} L.,    {Pelletier}
  G.,  2006, \aap, 447, 813

\bibitem[\protect\citeauthoryear{{Hannikainen}, {Campbell-Wilson}, {Hunstead},
  {McIntyre}, {Lovell}, {Reynolds}, {Tzioumis} \& {Wu}}{{Hannikainen}
  et~al.}{2001}]{han01}
{Hannikainen} D.,  {Campbell-Wilson} D.,  {Hunstead} R.,  {McIntyre} V.,
  {Lovell} J.,  {Reynolds} J.,  {Tzioumis} T.,    {Wu} K.,  2001, Astrophysics
  and Space Science Supplement, 276, 45

\bibitem[\protect\citeauthoryear{{Homan} \& {Belloni}}{{Homan} \&
  {Belloni}}{2005}]{hom05}
{Homan} J.,  {Belloni} T.,  2005, \apss, 300, 107

\bibitem[\protect\citeauthoryear{{King}, {Pringle}, {West} \& {Livio}}{{King}
  et~al.}{2004}]{kin04}
{King} A.~R.,  {Pringle} J.~E.,  {West} R.~G.,    {Livio} M.,  2004, \mnras,
  348, 111

\bibitem[\protect\citeauthoryear{{Lasota}}{{Lasota}}{2001}]{las01}
{Lasota} J.-P.,  2001, New Astronomy Review, 45, 449

\bibitem[\protect\citeauthoryear{{Lesur} \& {Longaretti}}{{Lesur} \&
  {Longaretti}}{2007}]{les07}
{Lesur} G.,  {Longaretti} P.-Y.,  2007, \mnras, 378, 1471

\bibitem[\protect\citeauthoryear{Lubow, Papaloizou, \& 
Pringle}{1994}]{lub94} Lubow S.~H., Papaloizou J.~C.~B., 
Pringle J.~E., 1994, \mnras, 267, 235 

\bibitem[\protect\citeauthoryear{{Maccarone} \& {Coppi}}{{Maccarone} \&
  {Coppi}}{2003}]{mac03c}
{Maccarone} T.~J.,  {Coppi} P.~S.,  2003, \mnras, 338, 189

\bibitem[\protect\citeauthoryear{{Meyer-Hofmeister}, {Liu} \&
  {Meyer}}{{Meyer-Hofmeister} et~al.}{2005}]{mey05}
{Meyer-Hofmeister} E.,  {Liu} B.~F.,    {Meyer} F.,  2005, \aap, 432, 181

\bibitem[\protect\citeauthoryear{{Miyamoto}, {Kitamoto}, {Hayashida} \&
  {Egoshi}}{{Miyamoto} et~al.}{1995}]{miy95}
{Miyamoto} S.,  {Kitamoto} S.,  {Hayashida} K.,    {Egoshi} W.,  1995, \apjl,
  442, L13

\bibitem[\protect\citeauthoryear{{Nespoli}, {Belloni}, {Homan}, {Miller},
  {Lewin}, {M{\'e}ndez} \& {van der Klis}}{{Nespoli} et~al.}{2003}]{nesp03}
{Nespoli} E.,  {Belloni} T.,  {Homan} J.,  {Miller} J.~M.,  {Lewin} W.~H.~G.,
  {M{\'e}ndez} M.,    {van der Klis} M.,  2003, \aap, 412, 235

\bibitem[\protect\citeauthoryear{{Nowak}, {Wilms} \& {Dove}}{{Nowak}
  et~al.}{2002}]{now02}
{Nowak} M.~A.,  {Wilms} J.,    {Dove} J.~B.,  2002, \mnras, 332, 856

\bibitem[\protect\citeauthoryear{{Pandey}, {Rao}, {Pooley}, {Durouchoux},
  {Manchanda} \& {Ishwara-Chandra}}{{Pandey} et~al.}{2006}]{pan06}
{Pandey} M.,  {Rao} A.~P.,  {Pooley} G.~G.,  {Durouchoux} P.,  {Manchanda}
  R.~K.,    {Ishwara-Chandra} C.~H.,  2006, \aap, 447, 525

\bibitem[\protect\citeauthoryear{{Petrucci}, {Ferreira}, {Cabanac}, {Henri} \&
  {Pelletier}}{{Petrucci} et~al.}{2006}]{pet06b}
{Petrucci} P.-O.,  {Ferreira} J.,  {Cabanac} C.,  {Henri} G.,    {Pelletier}
  G.,  2006, in Proceedings of the VI Microquasar Workshop. September 18-22,
  2006, Como, Italy.

\bibitem[\protect\citeauthoryear{{Remillard} \& {McClintock}}{{Remillard} \&
  {McClintock}}{2006}]{rem06}
{Remillard} R.~A.,  {McClintock} J.~E.,  2006, \araa, 44, 49

\bibitem[\protect\citeauthoryear{{Rodriguez}, {Corbel} \&
  {Tomsick}}{{Rodriguez} et~al.}{2003}]{rod03}
{Rodriguez} J.,  {Corbel} S.,    {Tomsick} J.~A.,  2003, \apj, 595, 1032

\bibitem[\protect\citeauthoryear{{Shakura} \& {Sunyaev}}{{Shakura} \&
  {Sunyaev}}{1973}]{sha73}
{Shakura} N.~I.,  {Sunyaev} R.~A.,  1973, \aap, 24, 337

\bibitem[\protect\citeauthoryear{{Smith}, {Heindl} \& {Swank}}{{Smith}
  et~al.}{2002}]{smi02}
{Smith} D.~M.,  {Heindl} W.~A.,    {Swank} J.~H.,  2002, \apj, 569, 362

\bibitem[\protect\citeauthoryear{{Sobczak}, {McClintock}, {Remillard}, {Cui},
  {Levine}, {Morgan}, {Orosz} \& {Bailyn}}{{Sobczak} et~al.}{2000}]{sob00}
{Sobczak} G.~J.,  {McClintock} J.~E.,  {Remillard} R.~A.,  {Cui} W.,  {Levine}
  A.~M.,  {Morgan} E.~H.,  {Orosz} J.~A.,    {Bailyn} C.~D.,  2000, \apj, 544,
  993

\bibitem[\protect\citeauthoryear{{Tagger}, {Varni{\`e}re}, {Rodriguez} \&
  {Pellat}}{{Tagger} et~al.}{2004}]{tag04}
{Tagger} M.,  {Varni{\`e}re} P.,  {Rodriguez} J.,    {Pellat} R.,  2004, \apj,
  607, 410

\bibitem[\protect\citeauthoryear{{Zanni}, {Ferrari}, {Rosner}, {Bodo} \&
  {Massaglia}}{{Zanni} et~al.}{2007}]{zan07}
{Zanni} C.,  {Ferrari} A.,  {Rosner} R.,  {Bodo} G.,    {Massaglia} S.,  2007,
  \aap, 469, 811

\bibitem[\protect\citeauthoryear{{Zdziarski} \& {Gierli{\'n}ski}}{{Zdziarski}
  \& {Gierli{\'n}ski}}{2004}]{zdz04}
{Zdziarski} A.~A.,  {Gierli{\'n}ski} M.,  2004, Progress of Theoretical Physics
  Supplement, 155, 99

\bibitem[\protect\citeauthoryear{{Zdziarski}, {Gierli{\'n}ski},
  {Miko{\l}ajewska}, {Wardzi{\'n}ski}, {Smith}, {Alan Harmon} \&
  {Kitamoto}}{{Zdziarski} et~al.}{2004}]{zdz04b}
{Zdziarski} A.~A.,  {Gierli{\'n}ski} M.,  {Miko{\l}ajewska} J.,
  {Wardzi{\'n}ski} G.,  {Smith} D.~M.,  {Alan Harmon} B.,    {Kitamoto} S.,
  2004, \mnras, 351, 791

\bibitem[\protect\citeauthoryear{{Zdziarski}, {Poutanen}, {Paciesas} \&
  {Wen}}{{Zdziarski} et~al.}{2002}]{zdz02b}
{Zdziarski} A.~A.,  {Poutanen} J.,  {Paciesas} W.~S.,    {Wen} L.,  2002, \apj,
  578, 357

\end{thebibliography}

\end{document}